# Massive coherent equipartition of light by the geometric phase of null space

Xiangrui Hou[1,*], Dongyi Wang[2,*], Fangyu Wang[1], Congwei Lu[2], Zhaoju Yang[1,†], Guancong Ma[2,3,†]

[1]*School of Physics and Zhejiang Key Laboratory of Micro-nano Quantum Chips and Quantum Control, Zhejiang University, Hangzhou 310058, Zhejiang Province, China*

[2]*Department of Physics, Hong Kong Baptist University, Kowloon Tong, Hong Kong, China*

[3]*Shenzhen Institute for Research and Continuing Education, Hong Kong Baptist University, Shenzhen 518000, China*

**Abstract**

Light source is a foundational to photonic science and technology. However, a significant challenge remains in generating and distributing coherent light from a single on-chip source with high phase stability across multiple channels. Integrated lasers typically operate independently, and conventional splitters (e.g., multi-mode interferometers) do not guarantee the phase coherence required for advanced applications. Here, we report a purely geometric scheme for achieving massive equipartition of coherent light on a photonic chip by leveraging the geometric phases of a null space spanned by degenerate states with zero eigenvalue. The evolution of the null space maps to real-space rotation described by the special orthogonal group $\mathrm{SO}(N)$, thus enabling precise and scalable control over light distribution by engineering the system parameters. We experimentally realize up to one-to-nine equipartition of light on a waveguide array fabricated on a glass-based photonic chip. The framework can be upscaled for one-to-$N$ light distribution. This work establishes a versatile and scalable platform for integrated coherent light sources, paving the way for integrated photonic applications such as quantum photonics and optical computing.

*These authors contributed equally to this work.
†Email: zhaojuyang@zju.edu.cn (Z. Y.)  phgcma@hkbu.edu.hk (G. M.)

**Introduction**

Coherent light source is an essential component in a large variety of optical and photonic applications such as communications [1-6], sensing [7-10], and computing [11-18]. With the ever-improving sophistication and demand for compactness, chip-integrable light sources are becoming increasingly important [19-24]. Integrated light sources, such as InP/AlGaAs lasers on silicon-based chips [19, 22, 25-29] and quantum-dot lasers [21, 30-33], are fabricated individually and operate in isolation. So there exist random phase factors among the emissions from different sources. This may compromise functionalities that require phase coherence, such as quantum optics and quantum computing [11-14, 34-35]. A viable route to overcome this challenge is to split the light from a single source. However, existing technologies such as multi-mode interferometers (MMIs) [25, 36-41] and directional couplers [41-45] lack inherent phase stability. In this work, we propose a new scheme for the coherent and one-to-many equipartition of light from a single source. The scheme hinges on the adiabatic evolution of the null space (or kernel) of a chiral-symmetric Hamiltonian, i.e., the subspace spanned by eigenmodes degenerate at zero eigenvalue. Under the further constraint of parity-time symmetry, the evolution of the null space is captured by an orthogonal matrix with a unity determinant [SO($N$), with $N$ being the dimension of the null space], which is the multi-state generation of geometric phase. The scheme is purely geometrical, and because all modes in the null space are degenerate, their coherence is not contaminated by the dynamic phase. The scheme is compatible with existing technologies of photonic chips. We implement this scheme using an array of waveguides fabricated in a low-loss glass platform using direct laser writing [46]. Up to one-to-nine splitting is experimentally achieved, and further upscaling is straightforward. As such, our work paves the way for on-chip quantum-optical applications in which coherent light sources are foundational.

**Results**

The principle of our scheme rests on the properties of a system described by a real-valued, chiral-symmetric Hamiltonian. Consider a $(M + N)$-dim Hamiltonian, with $N > M$,

$$H_{M,N}(\lambda) = \begin{pmatrix} 0_M & t^{\mathrm{T}}(\lambda) \\ t(\lambda) & 0_N \end{pmatrix}, \tag{1}$$

where $\lambda$ is a controlling parameter, $t(\lambda) \in \mathbb{R}^N \times \mathbb{R}^M$ is a full-rank $N \times M$ real matrix. It is straightforward to see that $\mathrm{rank}(H) = 2M$. From the rank-nullity theorem, $H$ always has an $(N - M)$- dim null space, denoted $\ker H$, i.e., a linear subspace of $H$ spanned by eigenmodes with a zero eigenvalue. In a physics sense, Eq. (1) exhibits an effective chiral symmetry $\Gamma^{-1}H\Gamma = -H$, with the chiral operator $\Gamma = I_M \oplus (-I_N)$, which protects $N - M$ degenerate zero modes. When the system is driven adiabatically by $\lambda$, the evolution of all $N - M$ zero modes in $\ker H$ produces diverse linear combinations described by

$$|\psi(\lambda)\rangle = \mathcal{U}|\psi(\lambda_0)\rangle, \tag{2}$$

where $|\psi(\lambda)\rangle$ and $|\psi(\lambda_0)\rangle$ are the end-state and initial-state vectors at zero energy, respectively. $\mathcal{U} = \exp[-i\int_\Gamma H(\lambda')d\lambda']$ is a unitary evolution operator. According to the adiabatic theorem, $\mathcal{U}$ contributes both dynamic and geometric effects. Since we are only concerned with modes at zero energy, $\mathcal{U}$ generates no dynamic phases, so its effect is purely geometric. We can thus write $\mathcal{U} = \mathcal{P}\exp[-\int_\Gamma A(\lambda')d\lambda']$, where $A(\lambda')$ is the Berry connection matrix with $A_{mn} = i\langle u_m(\lambda')|\partial_{\lambda'} u_n(\lambda')\rangle$ as entries, and $|u_{m,n}\rangle$ is the $m,n$-th zero-mode eigenvector under a chosen basis. $\Gamma$ denotes the evolution path, and $\mathcal{P}$ denotes path-ordering. Equation (2) then is recast into

$$|\psi(\lambda)\rangle = \sum_{m=1}^{N-M} U_{mn}|u_n(\lambda_0)\rangle, \tag{3}$$

where $U_{mn}$ is entries of $\mathcal{U}$.

Because $H = H^*$ and $H$ possesses time-reversal symmetry $K^{-1}HK = H^*$ with $K$ taking complex conjugation, all its eigenvectors are real vectors, i.e., $|u_n(\lambda)\rangle \in \mathbb{R}^{M+N}$. Therefore, the Berry connection is a traceless real antisymmetric matrix, which implies that the evolution of the zero modes is described by the special orthogonal (SO) group, i.e., $\mathcal{U} \in \text{SO}(N - M)$. It follows that the modes in $\ker H$ effectively undergo a real rotation in the adiabatic evolution.

To concretely describe the consequence of the above theory, let us consider a simplest case with $M = 1, N = 3$, such that

$$H_{1,3} = \begin{pmatrix} 0 & t_1 & t_2 & t_3 \\ t_1 & 0 & 0 & 0 \\ t_2 & 0 & 0 & 0 \\ t_3 & 0 & 0 & 0 \end{pmatrix}. \tag{4}$$

There are $N - M = 2$ zero modes in Hamiltonian (4). For convenience, we further set $t_1 = \cos\phi\sin\theta, t_2 = \sin\phi\sin\theta, t_3 = \cos\theta$, such that the evolution of $H$ in the $t_1 t_2 t_3$-space lies on a unit 2-sphere. It follows that the evolution of the two zero modes is captured by $\mathcal{U}_2(\Omega) = \begin{pmatrix} \cos\Omega & -\sin\Omega \\ \sin\Omega & \cos\Omega \end{pmatrix} \in \text{SO}(2)$, where $\Omega = \oint_\Gamma \sin\theta\, d\theta d\phi$ is the solid angle enclosed by a close path on the 2-sphere. It is observable that the final state of the evolution is determined by $\Omega$. When $\Omega = \pi/2$, $\mathcal{U}_2\left(\frac{\pi}{2}\right) = \begin{pmatrix} 0 & -1 \\ 1 & 0 \end{pmatrix}$, which swaps the two eigenvectors and a phase change of $\pi$ appears for the second mode. $\mathcal{U}_2\left(\frac{\pi}{2}\right)$ effectively braids the two zero modes – a phenomenon previously observed in acoustic and photonic waveguides [47-49].

It is readily clear that different output states can be obtained by choosing different $\Omega$. When $\Omega = \pi/4$, the evolution operator $\mathcal{U}_2\left(\frac{\pi}{4}\right) = \frac{1}{\sqrt{2}}\begin{pmatrix} 1 & -1 \\ 1 & 1 \end{pmatrix}$. An initial state of $|\psi(0)\rangle = (0, 1, 0, 0)^T$ evolves into $|\psi(L)\rangle = \frac{1}{\sqrt{2}}(0, 1, 1, 0)^T$ at the output. Here, the set of bases is $|u_1\rangle = (0, 1, 0, 0)^T$ and $|u_2\rangle = (0, 0, 1, 0)^T$, which are the eigenvectors of the initial Hamiltonian $H(\theta = 0)$. In other words, the energy initially dwelling in the second site is equally split between the second and the third sites, as shown in Fig. 1(a). Because of their zero eigenvalues, the states accumulate no dynamic phase during the splitting, so the equipartitioned waves at the two sites are coherently in phase.

Such equipartition can be realized in photonic chips. We fabricated a set of four evanescently coupled waveguides on Corning 7980 using the femtosecond laser direct-writing technique (see details in Supplementary Information, Section 3) [46], the schematic is shown in Fig. 1(d). The fundamental guiding mode plays the role of the onsite orbital in Eq. (4). Under the paraxial approximation [50], the dynamics of light along the propagation direction, denoted $z$, follows a Schrödinger-like equation $H(z)|\psi(z)\rangle = i\partial_z |\psi(z)\rangle$. The coupling coefficients $t_1, t_2, t_3$, which are embedded on a unit 2-sphere in the parameter space, are modulated as functions of $z$ according to the profiles shown in Fig. 1(b&c). This is achieved by varying the center-to-center spacing between waveguides from 8 to 24 μm, resulting in a variation of the normalized coupling coefficients between 0.03 and 1 for an injection of the 635-nm laser (see details in Supplementary Information, Section 4 & 5). We set the parametric loop to enclose a solid angle of $\Omega = \pi/4$, such that the light injected into waveguide 2 is pumped to waveguides 2 and 3 with a 50-50 distribution of energy. We successfully achieved such an equipartition of energy in our photonic experiments, as depicted in the results in Fig. 1(e). A similar phenomenon can be observed when injecting the light into waveguide 3, but the waves in the two sites are coherently out-of-phase (see details in Supplementary Information, Section 6&7), as shown in Fig. 1(f).

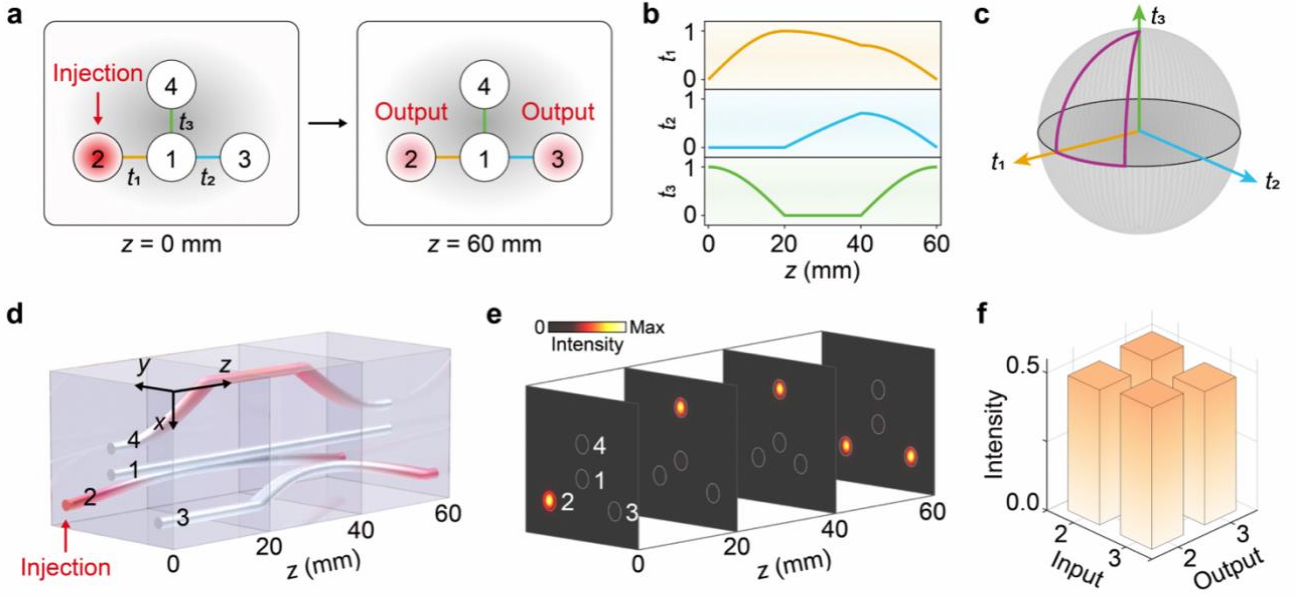

**Fig. 1 Equipartition of photonic energy by geometric phase of zero modes and its realization in photonic waveguides.** (a) The four-site tight-binding model [Eq. (4)] with two zero modes. The geometric phase can tune the superposition of the wavefunctions, thereby controlling the energy distribution. (b) The profiles of the coupling coefficients among the waveguides. They form a closed loop on a parametric 2-sphere enclosing a solid angle of $\Omega = \pi/4$, as depicted in (c). (d) The design of the waveguide array for photonic equipartition, in which the red color plots denote the propagation of light in the waveguides. The total length of the device is 60 mm. (e) Experimentally measured distribution of the light intensity at 20 mm, 40 mm, and at the output at 60 mm. The red circle indicates the light injection position. (f) The normalized intensity distributions at the output plane indicate equipartition of light for injection into sites 2 and 3, respectively.

We can extend the principle to achieve one-to-many equipartition. Consider a Hamiltonian with $11 \times 11$ in dimension and 8 in rank, such that the null space is three-dimensional, i.e., spanned by three zero modes. The Hamiltonian of the system is denoted by

$$H_{4,7} = \begin{pmatrix} 0 & A^{\mathrm{T}} & B^{\mathrm{T}} & C^{\mathrm{T}} \\ A & 0 & 0 & 0 \\ B & 0 & 0 & 0 \\ C & 0 & 0 & 0 \end{pmatrix}, \tag{5}$$

with $A = s_1 \times (1\ \ 1\ \ 1\ \ 1), B = s_2 \times I_4, C = s_3 \times \begin{pmatrix} 1 & 1 & 0 & 0 \\ 0 & 0 & 1 & 1 \end{pmatrix}$, where $I_4$ is a $4 \times 4$ identity matrix, $s_1, s_2, s_3$ are three independent coupling coefficients. The tight-binding model is plotted in Fig. 2(a), with the coupling coefficients modulated according to the profile shown in Fig. 2(b). In this case, the wavefunctions of the zero modes only support in sites 5-11. Despite the increased dimensionality of Eq. (5), the evolution of the zero modes still follows the description of SO(3). By choosing $\mathcal{U}_3\left(\cos^{-1}\frac{1}{\sqrt{17}}\right)$, an injection at the center of the cluster (site 5) is split five ways, producing an output equally distributed at sites 5-9, as depicted in Fig. 2(c). This splitting scheme is also realized in our system, as shown in Fig. 2(d). The distribution of output intensities is benchmarked, with the variation is < 5.6 % [Fig. 2(e)], and the waves between two of them with a coherently in-phase (see details in Supplementary Information, Section 6).

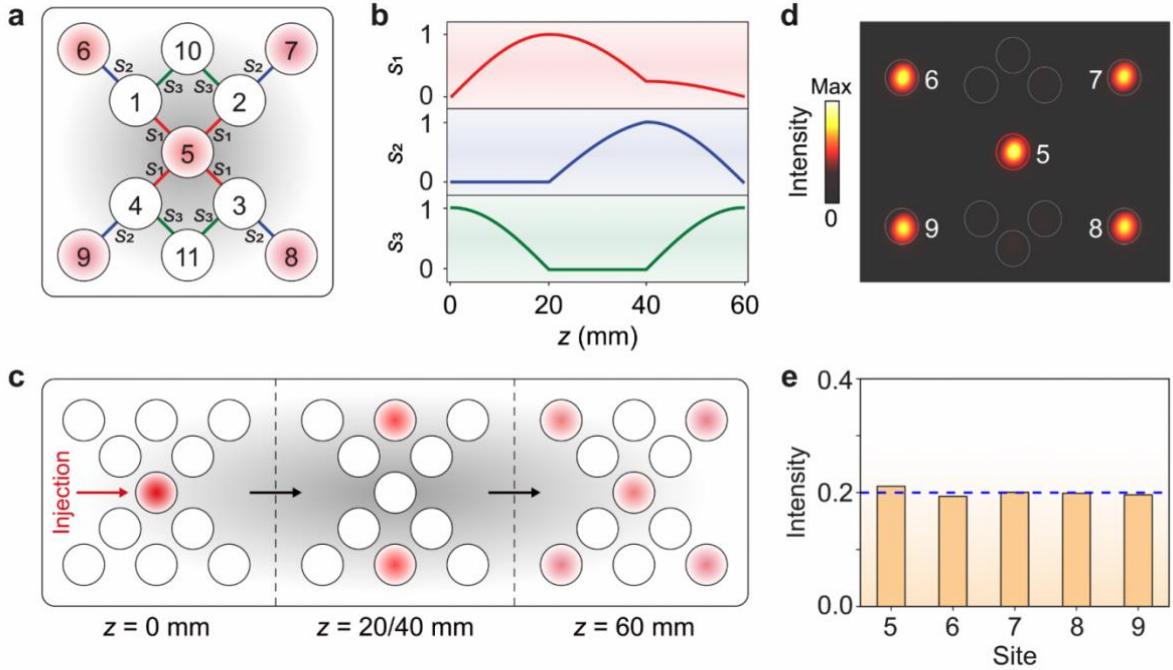

**Fig. 2  One-to-five photonic equipartition.** (a) The 11-site model sustaining three zero modes. (b) The variation of parameters along the propagation direction with a periodic length of 60 mm. (c) Computed wavefunction evolution when injecting at site 5. (d) Experimentally measured light intensities at the output. (e) The normalized distribution of energies at the outputs of waveguides 5-9.

The one-to-two and one-to-five equipartition designs can be cascaded to further increase the output count. The model shown in Fig. 3(a) combines a unit of one-to-five followed by four units of one-to-two to achieve one-to-nine splitting. Light is injected into the central waveguide (site 5), so the one-to-five splitting is executed first. Note that the SO(3) rotation in this stage must be adjusted to $\mathcal{U}_3\left(\cos^{-1}\frac{1}{\sqrt{33}}\right)$, such that the end state at sites 6-9 has twice the intensity compared to site 5. Then, the light at sites 6-9 each propagates through a unit of one-to-two equipartition, which is identical to the one described in Fig. 1. The parameter profiles along the propagation direction is shown in Fig. 3(b). This cascaded design produces a near-uniform intensity distribution across all nine output waveguides, as confirmed experimentally in Fig. 3(c, d).

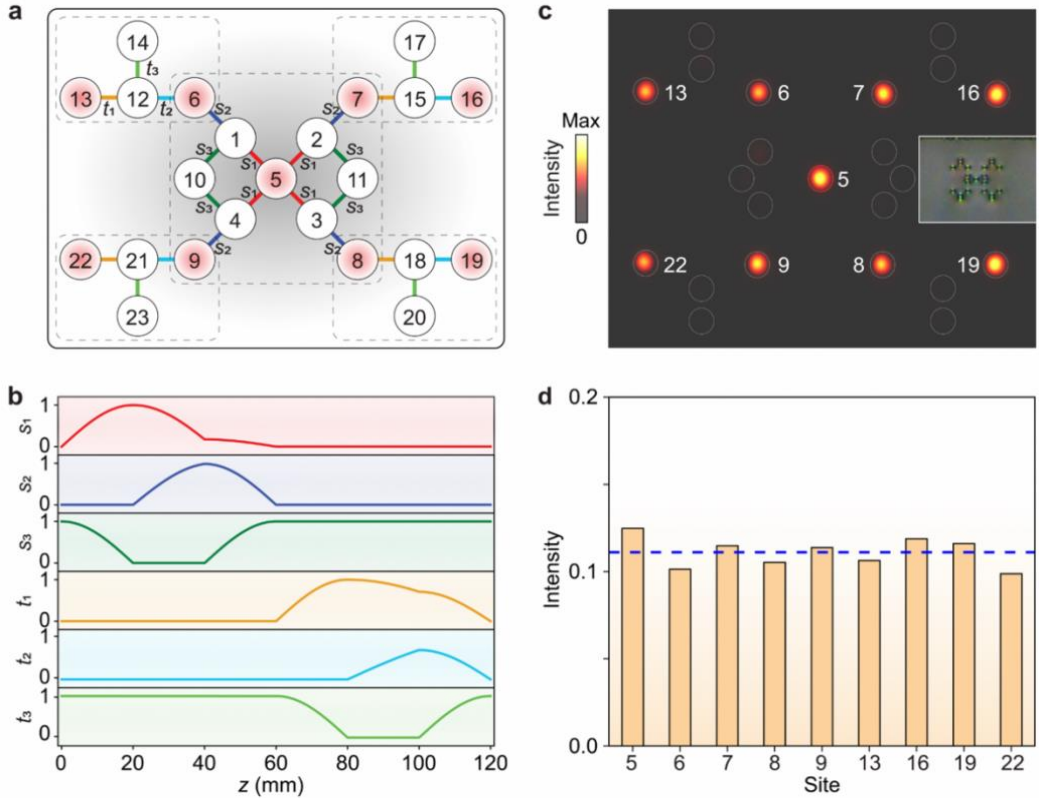

**Fig. 3 One-to-nine photonic equipartition.** (a) The model combines an 11-site cluster and four 4-site clusters. (b) The modulation profiles of the coupling coefficients among the waveguides. (c) Experimentally measured light intensities at the output show near-equal intensity at the nine designated waveguides. The inset is a photographic image of the cross-section of the waveguide array at the output. (d) The normalized distribution of intensities at the nine waveguides.

It is straightforward to further expand the split count by increasing the dimensionality of the null space. But this means that the set of zero modes undergoes high-dimensional rotation, which must be simultaneously controlled by an increasing number of parameters. An alternative approach to increase the split count is to sequentially compound different systems.

We now demonstrate the scalability of our approach by extending it to a 1-to-$N$ ($N = 41$) coherent equipartition scheme, as shown in Fig. 4. Light injected into the central site is evenly distributed across all red-colored sites at the output. The layout consists of three different types of partition designs: one unit of one-to-five (unequal) splitting at the center, twelve units of one-to-three splitting marked with yellow, and twelve units of one-to-two splitting marked with blue.

Next, we elaborate on how this design works. The 1-to-$N$ equipartition is divided into $n$ steps with $N = n^2 + (n+1)^2$. (Apparently, $n = 4$ in Fig. 4.) The splitting sequence begins with a central one-to-five splitting, which distributes light from the injection point into four symmetric sectors. And then the light is further split into these four identical sectors after the first step. So the subsequent $i$ steps ($n \geq i > 1$) consist of four one-to-three splitting and $(i-2)$ one-to-two splitting.

Now, we determine the distribution ratio at each step. We denote the distribution on the site $(i,j)$ before splitting as $\psi_{i,j}$, and the distribution after splitting as $\tilde{\psi}_{i,j}$. For the first step, the light experiences a one-to-five splitting, we have $(\psi_{n+1,1}, \psi_{n,1}, \psi_{n+1,0}, \psi_{n+2,1}, \psi_{n+1,2}) = (1,0,0,0,0)$, and $(\tilde{\psi}_{n+1,1}, \tilde{\psi}_{n,1}, \tilde{\psi}_{n+1,0}, \tilde{\psi}_{n+2,1}, \tilde{\psi}_{n+1,2}) = \frac{1}{\sqrt{N}}\left(1, \sqrt{\frac{(1+n)n}{2}}, \sqrt{\frac{(1+n)n}{2}}, \sqrt{\frac{(1+n)n}{2}}, \sqrt{\frac{(1+n)n}{2}}\right)$. For the one-to-three units, we have $(\psi_{i,1}, \psi_{i,2}, \psi_{i-1,1}) = \frac{1}{\sqrt{N}}\left(\sqrt{\frac{(1+i)i}{2}}, 0, 0\right)$, and $(\tilde{\psi}_{i,1}, \tilde{\psi}_{i,2}, \tilde{\psi}_{i-1,1}) = \frac{1}{\sqrt{N}}\left(1, \sqrt{i-1}, \sqrt{\frac{(i-1)i}{2}}\right)$, where $1 < i \leq n$. For the one-to-two units, $(\psi_{i,j}, \psi_{i+1,j}) = \frac{1}{\sqrt{N}}\left(\sqrt{i-j+1}, 0\right)$, and $(\tilde{\psi}_{i,j}, \tilde{\psi}_{i+1,j}) = \frac{1}{\sqrt{N}}\left(1, \sqrt{i-j}\right)$, where $2 < i \leq n$ and $1 < j < i$. It is noteworthy that, after each splitting operation, the distribution at the site where the incident light located becomes $1/\sqrt{N}$. This process results in an equal distribution of the incident light across all $N$ sites at the output.

Next, we determine the solid angles corresponding to each splitting operation. Consider the simplest case of one-to-two splitting. The distribution of the outgoing light at the two sites is $(\tilde{\psi}_{i,j}, \tilde{\psi}_{i+1,j}) = \frac{1}{\sqrt{N}}\left(1, \sqrt{i-j}\right) \propto (\cos \Omega^{12}, \sin \Omega^{12})$. Then, it is straightforward to work out $\Omega^{12} = \tan^{-1}\sqrt{i-j}$. For the one-to-three splitting, it is obvious that it can be break down into two overlapped one-to-two splitting units. Therefore, the final distribution can be expressed as a superposition of one-to-two splitting in two orthogonal directions, i.e., $(\tilde{\psi}_{i,1}, \tilde{\psi}_{i,2}, \tilde{\psi}_{i-1,1}) = \frac{1}{\sqrt{N}}\left(1, \sqrt{i-1}, \sqrt{\frac{(i-1)i}{2}}\right) \propto$ $(\cos \Omega_1^{13} + \cos \Omega_2^{13}, \sin \Omega_1^{13}, \sin \Omega_2^{13})$, where $\Omega_{1,2}^{13}$ are the solid angles of the one-to-two splitting in each direction. Then we have $\Omega_1^{13} = \sin^{-1}\left(\frac{4\sqrt{n-1}}{\sqrt{n}\sqrt{-8+17n-6n^2+n^3}}\right)$ and $\Omega_2^{13} = \sin^{-1}\left[\frac{2\sqrt{2}\sqrt{(-1+n)}}{\sqrt{-8+17n-6n^2+n^3}}\right]$. Similarly, we obtain that the one-to-five splitting shall follow $(\tilde{\psi}_{n+1,1}, \tilde{\psi}_{n,1}, \tilde{\psi}_{n+1,0}, \tilde{\psi}_{n+2,1}, \tilde{\psi}_{n+1,2}) \propto$ $(4\cos\Omega^{15}, \sin\Omega^{15}, \sin\Omega^{15}, \sin\Omega^{15}, \sin\Omega^{15})$, so $\Omega^{15} = \tan^{-1}\sqrt{8n(n+1)}$.

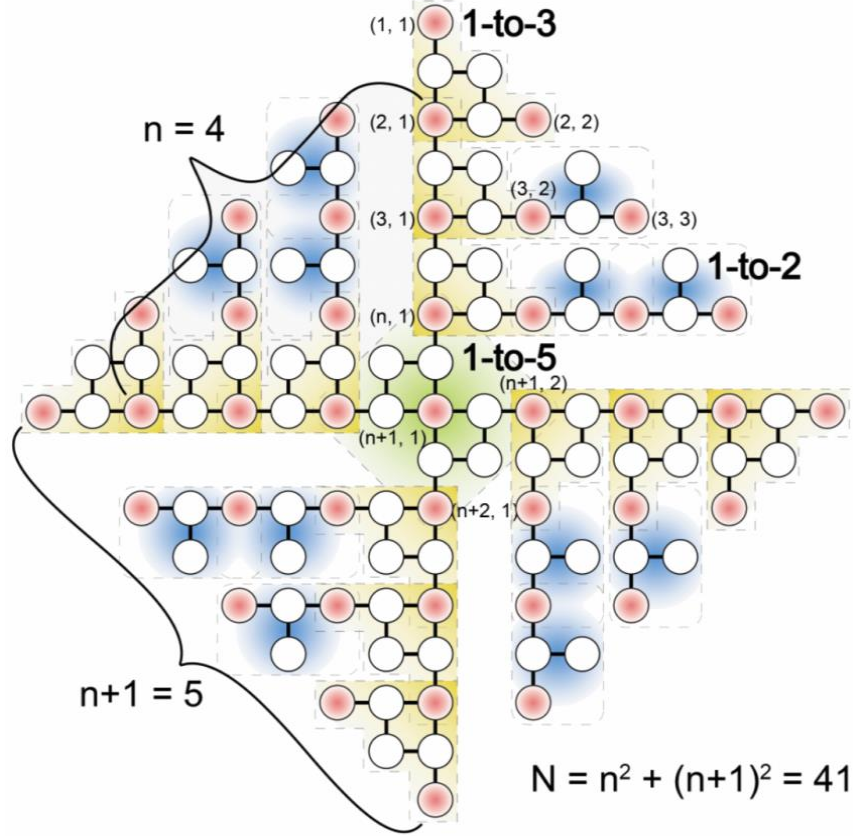

**Fig. 4. A scalable design for 1-to-$N$ light splitting with in-phase, equal intensity outputs.** The schematic depicts a 1-to-41 splitting design, demonstrating a modular and expandable massive coherent light splitting. The green, orange, and blue shades indicated 1-to-5, 1-to-3, and 1-to-2 splitting units.

**Conclusions**

In summary, we introduce an approach for achieving massive coherent light equipartition on photonic chips by exploiting the geometric phases of the null spaces of the chiral-symmetric Hamiltonians. Through designing the specific solid angles and controlling the adiabatic evolution of degenerate zero modes in specially designed waveguide arrays, we harness SO($N$) rotations to precisely control light distribution across multiple output channels. Experimentally, we demonstrate high-fidelity one-to-two, one-to-five, and one-to-nine splitting, confirming excellent phase stability and coherence. Our approach can also coherently generate staggered anti-phase light at the output, as shown in supplementary note 7. It is important to highlight that the proposed devices are reciprocal. This allows for inverse operation as coherent beam combiners, offering a direct route to integrated high-intensity light sources. The proposed modular architecture, based on cascading fundamental splitting units, enables scalable and versatile 1-to-$N$ light splitting, thereby providing a powerful platform for advanced integrated photonic applications.

We anticipate that the scheme reported in this paper will be important for next-generation integrated photonic applications with stringent requirements in optical coherence. These applications are likely associated with quantum light and light-based quantum computation. Future endeavours can include using photonic quantum metrics [51-52] to accelerate the adiabatic evolution, exploring non-adiabatic schemes [53-56] for rapid operations and reduced device footprints, integrating with active devices and exploring nonlinear phenomena [57, 58] to enhance functionality.

**Acknowledgements.** This work was supported by supported by the National Natural Science Foundation of China (NSFC, T2525002), the National Key R&D Program (2023YFA1406703, 2022YFA1404203, 2022YFA1404400), the Hong Kong Research Grants Council (RFS2223-2S01, 12301822, 12300925), the Hong Kong Baptist University (RC-RSRG/23-24/SCI/01, RC-SFCRG/23-24/R2/SCI/12), and the Fundamental Research Funds for the Central Universities (Grant No. 226-2025-00124).

# Supplementary Information

# Massive coherent equipartition of light by the geometric phase of null space

Xiangrui Hou[1,*], Dongyi Wang[2,*], Fangyu Wang[1], Congwei Lu[2], Zhaoju Yang[1,†], Guancong Ma[2,3,†]

[1]School of Physics and Zhejiang Key Laboratory of Micro-nano Quantum Chips and Quantum Control, Zhejiang University, Hangzhou 310058, Zhejiang Province, China

[2]Department of Physics, Hong Kong Baptist University, Kowloon Tong, Hong Kong, China

[3]Shenzhen Institute for Research and Continuing Education, Hong Kong Baptist University, Shenzhen 518000, China

## 1. Evolution of light in the photonic lattice

Here, we give a brief derivation of the paraxial propagation equation of light in the photonic lattice. Begin with the Maxwell's wave equation in a medium with refractive index of $n = n(x, y, z)$:

$$\nabla^2 \mathbf{E} + k_0^2 n^2 \mathbf{E} = 0, \qquad (S1)$$

where $\mathbf{E}$ is the electric field vector and $k_0 = \omega/c$ is the wavenumber. Rewriting equation (1) into the scalar form (Helmholtz equation), we get

$$\frac{\partial^2 E}{\partial x^2} + \frac{\partial^2 E}{\partial y^2} + \frac{\partial^2 E}{\partial z^2} + k_0^2 n^2 E = 0. \qquad (S2)$$

By inserting the solutions $E(x, y, z) = \psi(x, y, z) e^{i(k_0 n_0 z + \omega t)}$ into the equation (S2), we obtain

$$\nabla_\perp^2 \psi + 2i k_0 n_0 \frac{\partial \psi}{\partial z} + k_0^2 (n^2 - n_0^2) \psi = 0, \qquad (S3)$$

where $n_0 = n - \Delta n$ is the background refractive index of the medium, and $\Delta n$ is the refractive index change during the fabrication. Ignoring the quadratic term of $\Delta n$, we can make the approximation $(n^2 - n_0^2) \cong 2n_0(n - n_0) = 2n_0 \Delta n$. Then Eq. (S3) can be expressed as [1]

$$i \frac{\partial \psi}{\partial z} = -\left( \frac{1}{2 k_0 n_0} \nabla_\perp^2 - k_0 \Delta n \right) \psi, \qquad (S4)$$

which is the paraxial equation of light propagating in the photonic lattice. Equation S4 has the same mathematical form as the Schrödinger equation, thus the operator on the right-hand side is regarded as the Hamiltonian.

## 2. Determining the relative phase of two ports

To obtain the relative phase of light between the arbitrary two ports of the system, we can extend the waveguides of the two ports and introduce an additional waveguide located at the midpoint of the two waveguides. Consider the evolution of light in a three-waveguide system with equidistance:

$$-i\frac{\partial \psi_1}{\partial z} = \beta_1 \psi_1 + c\psi_2$$
$$-i\frac{\partial \psi_3}{\partial z} = \beta_3 \psi_3 + c\psi_2 \quad \text{(S5)}$$
$$-i\frac{\partial \psi_2}{\partial z} = \beta_2 \psi_2 + c\psi_1 + c\psi_3.$$

For the same propagation constant ($\beta_1 = \beta_2 = \beta_3 = \beta$), we can deduce the solutions of Eq. (S5), which are

$$\psi_1(z) = A_0 e^{i\beta z} + A_+ e^{i(\beta+\sqrt{2}c)z} + A_- e^{i(\beta-\sqrt{2}c)z}$$
$$\psi_2(z) = \sqrt{2}(A_+ e^{i(\beta+\sqrt{2}c)z} - A_- e^{i(\beta-\sqrt{2}c)z}) \quad \text{(S6)}$$
$$\psi_3(z) = -A_0 e^{i\beta z} + A_+ e^{i(\beta+\sqrt{2}c)z} + A_- e^{i(\beta-\sqrt{2}c)z},$$

where the coefficients $A_0 = [\psi_1(0) - \psi_3(0)]/2$, $A_+ = [\psi_1(0) + \psi_3(0) + \sqrt{2}\psi_2(0)]/4$, $A_- = [\psi_1(0) + \psi_3(0) - \sqrt{2}\psi_2(0)]/4$ and $\boldsymbol{\psi}(0) = [\psi_1(0), \psi_2(0), \psi_3(0)]^T$ is the initial condition.

In our system, the intensity of light emitted from any two ports is equal, and the relative phase of them is either in-phase or out-of-phase. So the initial condition of Eq. (S6) is $\boldsymbol{\psi}(0) = [1, 0, \pm 1]^T/\sqrt{2}$.

For the case of in-phase, the coefficients of Eq. (S6) are $A_0 = 0, A_+ = A_- = \sqrt{2}/4$, so we obtain

$$\psi_1(z) = (e^{i\sqrt{2}cz} + e^{-i\sqrt{2}cz})e^{i\beta z}/2\sqrt{2} = \cos(\sqrt{2}cz)e^{i\beta z}/\sqrt{2}$$
$$\psi_2(z) = (e^{i\sqrt{2}cz} - e^{-i\sqrt{2}cz})e^{i\beta z}/2 = i\sin(\sqrt{2}cz)e^{i\beta z} \quad \text{(S7)}$$
$$\psi_3(z) = (e^{i\sqrt{2}cz} + e^{-i\sqrt{2}cz})e^{i\beta z}/2\sqrt{2} = \cos(\sqrt{2}cz)e^{i\beta z}/\sqrt{2}.$$

The intensities of light in waveguide 1, 3 and waveguide 2 are $|\psi_{1,3}(z)|^2 = |\cos(\sqrt{2}cz)|^2/2$ and $|\psi_2(z)|^2 = |\sin(\sqrt{2}cz)|^2$, respectively. The oscillation period of the intensity evolution in the waveguides is $T = \pi/\sqrt{2}c$. The schematic of the waveguide array is shown in Fig. S1a, and the simulated result of the relationship between the light intensity of the waveguide and the propagation length is shown in Fig. S1b.

For the case of out-of-phase, the coefficients of equation (6) are $A_0 = \sqrt{2}/2$, $A_+ = A_- = 0$, so we obtain

$$\begin{aligned}\psi_1(z) &= \sqrt{2}e^{i\beta z}/2 \\ \psi_2(z) &= 0 \\ \psi_3(z) &= -\sqrt{2}e^{i\beta z}/2.\end{aligned} \quad (8)$$

In this case, the light intensity in waveguides 1 and 3 remains unchanged, and the intensity in waveguide 2 is always zero.

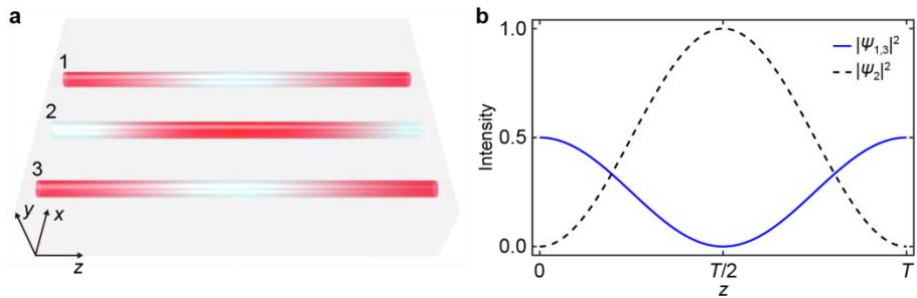

**Fig. S1 Interference in the three-waveguide system. a.** Schematic of the waveguide array. The red area denotes the light intensity in the waveguides. **b**, Relationship between the light intensity and the propagation length $z$.

### 3. Sample fabrication

The photonic waveguide systems for experimental characterization are fabricated on fused silica glass (Corning 7980) utilizing femtosecond laser writing [2] (repetition rate 600 kHz, center wavelength 515 nm, pulse duration 270 fs). The laser beam is focused into the sample through the objective lens (NA = 0.45), inducing the change of the refractive index up to $\Delta n = 0.7 \times 10^{-4}$ in the experiment. By modulating the displacement platform, thus controlling the position of the sample, curved-shaped waveguides can be arbitrarily fabricated.

### 4. Experimental setup

The experimental setup is shown in Fig. S2. The super continuous spectrum emitted by the laser first passes through a BPF (bandpass filter) to screen out the laser beam with a central wavelength of 635nm. Then the beam is focused onto the rotated mirror M$_2$, and then reflected to the input facet of the sample through a 4-f system (combination of lens $L_2$ & $L_3$). The result of the dynamic evolution of the beam is captured by the CCD after passing through the lens $L_4$.

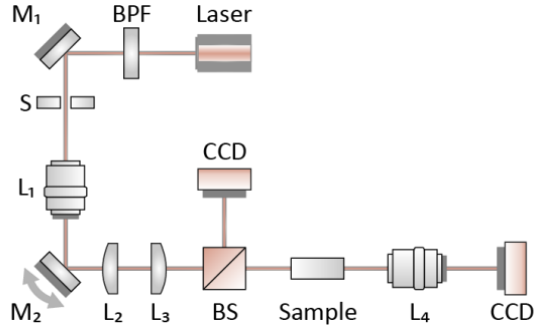

**Fig. S2.** The schematic diagram of the experimental setup.

## 5. Coupling strength on the distance between waveguides

In the experiment, the coupling strengths are tuned by the distance between two neighbouring waveguides. Fig. S3 plots the experimentally measured coupling strength as a function of the distance, together with a fitting of an exponential function. The positions of the waveguides are determined by referencing this benchmark.

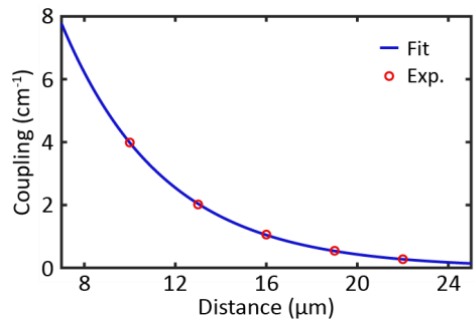

**Fig. S3.** The dependence of the coupling strength on the distance between two waveguides.

## 6. Measurement of the coherency of light at the output

Our light partition scheme is purely geometric and based on the degeneracy of the waveguide modes. Consequently, the light emitted at output ports is coherent. To validate this important characteristic, we use an interference experiment to measure the phases of the output light, the theoretical derivation is shown in Section 2.

Figure S4a illustrates the schematic used in the experiment to measure the phase of the one-to-two equipartition. An interference section is placed after the equipartition section, in which an additional waveguide 5 is placed in the middle of waveguides 2 and 3. The placement ensures equal and symmetric coupling to both adjacent waveguides, which enables the output light from waveguides 2 and 3 to be coherently superposed in waveguide 5.

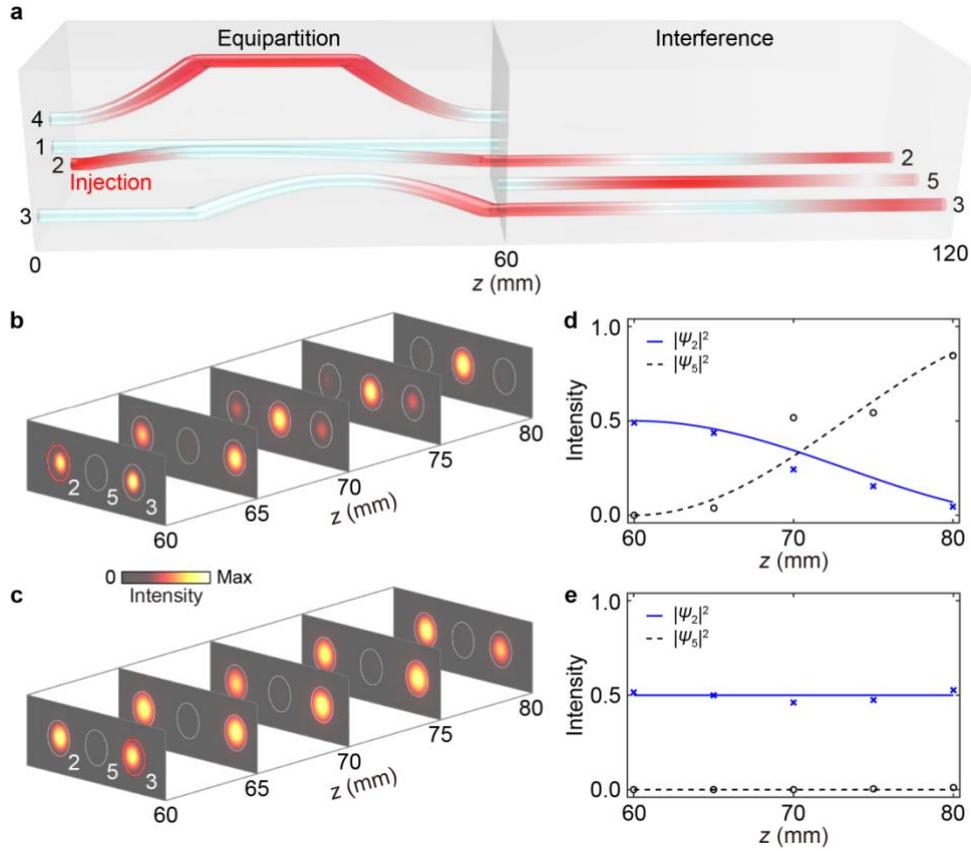

**Fig. S4 Measurement for the coherent phase of the one-to-two equipartition. a**, Schematic of the waveguide layout, in which the red color plots denote the propagation of light in the waveguides when injecting light into waveguide 2. The system contains an equipartition section (0–60 mm), followed by an interference section (60–80 mm). **b**, **c**, The experimentally measured intensity profiles at different propagation lengths under two distinct injections. **d**, **e**, The normalized intensity of the waveguides 2 and 5 at different propagation lengths. The coupling coefficient of waveguides 2 and 5 is about 0.42 cm$^{-1}$.

Fig. S4b-e shows the results of the interference. When light is injected into waveguide 2, constructive interference occurs in waveguide 5, leading to periodic intensity oscillations (i.e., a breathing-like behavior, see Fig. S4b&d, which means that waves in the two waveguides are in-phase.

Following the same methodology, we performed interference measurements on the phases of the output light of the one-to-five equipartition. Owing to the structural symmetry, we characterized the relative phase of the quadrant of the structure, i.e., waveguides 5 and 6, as shown in Fig. S5.

This methodology enables the determination of the coherent phase at any output port in the designed distribution network.

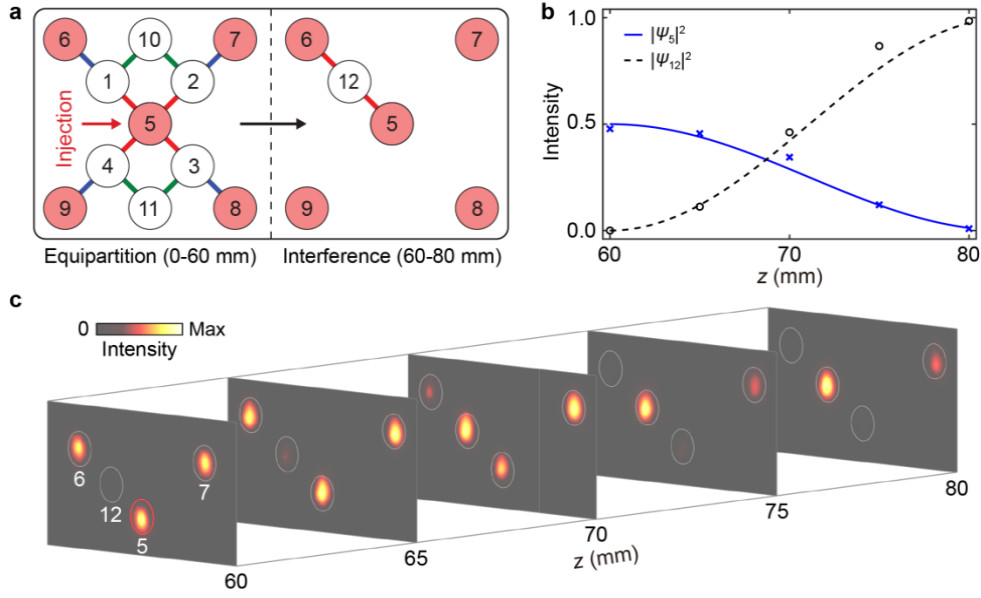

**Fig. S5 Measurement for the coherent phase of the one-to-five equipartition. a**, Tight-binding model of the system, which contains an equipartition section (0–60 mm), followed by an interference section (60–80 mm). **b**, The normalized intensity of the waveguides 5 and 12 at different propagation lengths. **c**, Experimentally measured optical intensity profiles at propagation lengths of $z = 65$ mm, 70 mm, 75 mm, and 80 mm, respectively (showing the upper portion for clarity). The coupling coefficient of waveguides 5 and 12 is about 0.5 cm$^{-1}$.

## 7. Anti-phase energy equipartitions

Our scheme can achieve anti-phase energy equipartition. For one-to-two equipartition, a simple way is to use the same structure as the one-to-two partition in the main text, i.e., $\mathcal{U}_2\left(\frac{\pi}{4}\right) = \frac{1}{\sqrt{2}}\begin{pmatrix} 1 & -1 \\ 1 & 1 \end{pmatrix}$, but a different initial state as $|\psi(0)\rangle = |u_2\rangle = (0, 0, 1, 0)$, i.e., injection at site 3. The resulting output state $|\psi(L)\rangle = \frac{1}{\sqrt{2}}(0, -1, 1, 0)^{\mathrm{T}}$ is in anti-phase. The evolution of the coupling parameters and the state distribution are shown in Fig. S6a and b, respectively.

If we keep the same initial state, i.e., $|\psi(0)\rangle = |u_2\rangle = (0, 1, 0, 0)$, then the evolution profile for achieving anti-phase output is $\mathcal{U}_2\left(\frac{\pi}{4}\right)\mathcal{U}_2\left(\frac{\pi}{2}\right) = \frac{1}{\sqrt{2}}\begin{pmatrix} 1 & 1 \\ -1 & 1 \end{pmatrix}$. In the first step, we follow the evolution path described in the main text to obtain the state $|\psi(L_1)\rangle = \frac{1}{\sqrt{2}}(0, 1, 1, 0)^{\mathrm{T}}$. In the second step, we evolve the coupling parameters along a path that subtends a solid angle of $\pi/2$, with the corresponding evolution operator $\mathcal{U}_2\left(\frac{\pi}{2}\right) = \begin{pmatrix} 0 & -1 \\ 1 & 0 \end{pmatrix}$, obtaining the anti-phase state $|\psi(L_2)\rangle = \frac{1}{\sqrt{2}}(0, -1, 1, 0)^{\mathrm{T}}$. This is shown in Fig. S6c and d, respectively. The upper and lower loops on the right side of Fig. S6d represent the parameter evolution paths of the first and second steps on the two spheres, respectively.

For one-to-five equipartition, we set $\Omega = \pi - \cos^{-1}\frac{1}{\sqrt{17}}$. An injection at the center of the cluster (site 5) is split five ways, producing an output equally distributed at sites 5-9, the phase in site 5 is opposite to that in sites 6–9, as depicted in Fig. S7b. The evolution of the coupling parameters is shown in Fig. S7a.

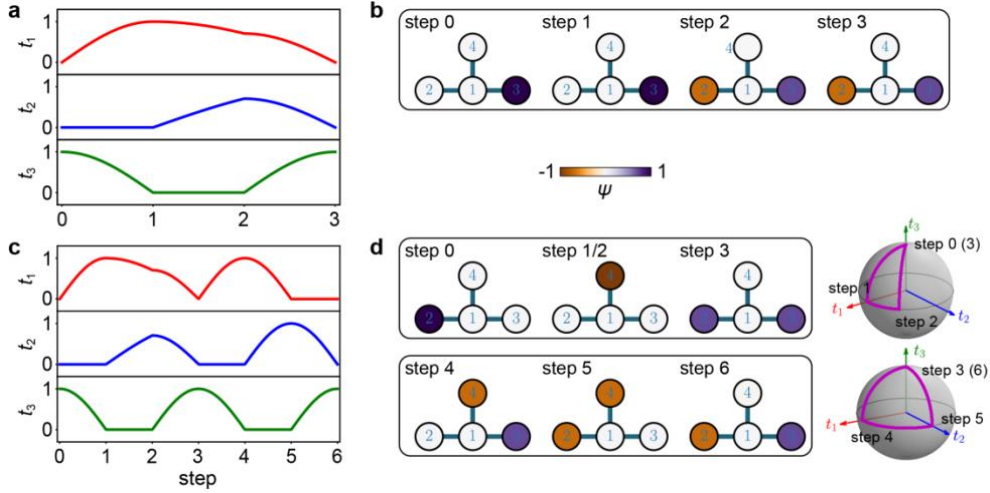

**Fig. S6 One-to-two anti-phase energy equipartition. a**, The profiles of the coupling coefficients for the first scheme. **b**, The distribution of energies at waveguides for the first scheme. **c**, The profiles of the coupling coefficients for the second scheme. **d**, The evolution of distribution for the first step (upper panel) and second step (lower panel).

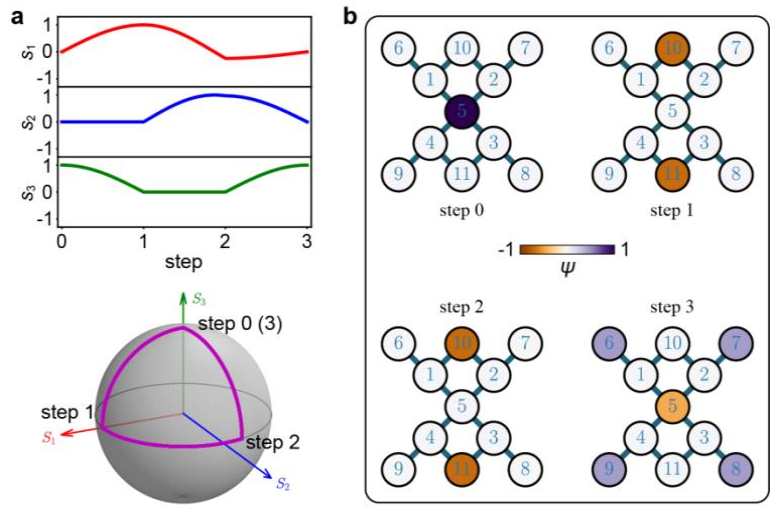

**Fig. S7 One-to-five anti-phase energy equipartition. a**, The profiles of the coupling coefficients. **b**, The distribution of energies at waveguides.